\documentclass[prd,aps,twocolumn,a4paper,floatfix,nofootinbib]{revtex4-1}

\usepackage[utf8]{inputenc}
\usepackage{graphicx,psfrag}
\usepackage{mathrsfs}
\usepackage{amsmath,amsfonts,amssymb}
\usepackage{multirow}
\usepackage{comment}
\usepackage{xcolor}
\usepackage{enumerate}
\usepackage{hyperref}
\hypersetup{
    colorlinks = true,
    linkcolor = {blue},
    citecolor = {blue},
    urlcolor = {blue},
    linkbordercolor = {white},
    }
\usepackage{caption}
\usepackage{subcaption}

\newcommand{\be}{\begin{equation}}
\newcommand{\ee}{\end{equation}}
\newcommand{\bea}{\begin{eqnarray}}
\newcommand{\eea}{\end{eqnarray}}
\newcommand{\bel}{\begin{align}}
\newcommand{\eel}{\end{align}}
\newcommand{\bi}{\begin{itemize}}
\newcommand{\ei}{\end{itemize}}

\def\non{\nonumber}                     
\def\half{\frac{1}{2}}

\def\Msun{{\rm M_{\odot}}}
\def\GMc2{{\rm G M_{\odot} c^{-2}}}

\def\bam{\texttt{BAM}}
\def\lorene{\texttt{LORENE}}
\def\z4c{Z4c}

\usepackage{color}

\definecolor{cyan}{rgb}{0,0.9,0.9}
\definecolor{orange}{rgb}{0.9,0.5,0}
\definecolor{magenta}{rgb}{1,0,1}
\definecolor{purple}{rgb}{0.8,0.4,0.8}
\definecolor{gray}{rgb}{0.8242,0.8242,0.8242}
\definecolor{mgreen}{rgb}{0.1,0.8,0.1}

\usepackage[normalem]{ulem}

\begin{document}

\title{Black-Hole Neutron Star Simulations with the \bam~code: 
First Tests and Simulations}

\author{Swami Vivekanandji \surname{Chaurasia}$^{1}$}
\author{Tim \surname{Dietrich}$^{2,3}$}
\author{Stephan \surname{Rosswog}$^{1}$}

\affiliation{${}^1$The Oskar Klein Centre, Department of Astronomy, Stockholm University, AlbaNova, SE-10691 Stockholm,
Sweden}
\affiliation{${}^2$Institut f\"{u}r Physik und Astronomie, Universit\"{a}t Potsdam, Haus 28, Karl-Liebknecht-Strasse 24/25, 14476, Potsdam, Germany}
\affiliation{${}^3$Max Planck Institute for Gravitational Physics (Albert Einstein Institute),
Am M\"{u}hlenberg 1, Potsdam 14476, Germany}

\date{\today}

\begin{abstract}
The first detections of black hole - neutron star mergers (GW200105 and
GW200115) by  the LIGO-Virgo-Kagra Collaboration mark a significant scientific
breakthrough. The physical interpretation of pre- and post-merger signals requires 
careful cross-examination between observational and theoretical modelling results.
Here we present the first set of black hole - neutron star simulations that were
obtained with the numerical-relativity code \bam. Our initial data are constructed
using the public \lorene\ spectral library which employs an excision of the black hole interior.
\bam{}, in contrast, uses the moving-puncture gauge for the evolution.
Therefore, we need to ``stuff'' the black hole interior with smooth initial
data to evolve the binary system in time. This procedure introduces 
constraint violations such that the constraint damping properties of the evolution 
system are essential to increase the accuracy of the simulation and  in particular to 
reduce spurious center-of-mass drifts.
Within \bam\ we evolve the \z4c equations and we compare
our gravitational-wave results with those of the SXS collaboration and results
obtained with the SACRA code. While we find generally good agreement with
the reference solutions and phase differences $\lesssim 0.5$~rad at the moment of 
merger, the absence of a clean convergence order in our simulations does not allow
for a proper error quantification. We finally present a set of different initial conditions
to explore how the merger of black hole neutron star systems depends on the involved 
masses, spins, and equations of state.
\end{abstract}

\pacs{
  04.25.D-,     
  04.30.Db,   
  95.30.Sf,     
  95.30.Lz,   
  97.60.Jd      
  97.60.Lf    
  98.62.Mw    
}

\maketitle

\section{Introduction}
\label{sec:intro}

With the first direct detection of the gravitational waves (GWs) \cite{Abbott:2016blz}
from a merging black hole binary GWs have become an active part of observational astronomy with a constantly increasing number of detected events ~\cite{LIGOScientific:2018mvr,Abbott:2020niy}. After binary black hole (BBH) and
binary neutron star (BNS) mergers~\cite{Abbott:2017BNS} most recently the
detection of GW200105 and GW200115 \cite{LIGOScientific:2021qlt} 
has also demonstrated the existence of ``mixed binaries'' that consist of a black hole orbited by a neutron star (BHNS).

The majority of GW signals detected so far comes 
from BBH mergers. Only two BNS systems have been seen, 
namely GW170817~\cite{Abbott:2018wiz} and GW190425~\cite{Abbott:2020uma}. 
While there is clear evidence that these systems have been BNSs, 
either through electromagnetic signals or based on binary population studies, 
the GW signal alone would not be sufficient to distinguish GW170817 
and GW190425 from BHNS mergers, e.g., 
\cite{Coughlin:2019kqf,Hinderer:2018pei,Kyutoku:2020xka}.
Similarly, also GW190814~\cite{Abbott:2020khf}, which has been very likely 
a BBH merger, e.g.,~\cite{Essick:2020ghc,Tews:2020ylw}, could have been a BHNS system.

In addition, the observation of GW200105 and GW200115 mark the first confirmed detections 
of BHNS systems. Based on the large mass ratio and small BH spin, it was expected that these systems will not produce bright electromagnetic counterpart, hence, the non-detection of electromagnetic signatures comes as no surprise. However, based on the GW signal alone, it was possible to constrain the component masses of GW200105 to be $8.9^{+1.2}_{-1.5}M_\odot$ and $1.9^{+0.3}_{-0.2}M_\odot$, and for GW200115 to be $5.7^{+1.8}_{-2.1}M_\odot$ and $1.5^{+0.7}_{-0.3}M\odot$ (at the 90\% credible level).

To extract such information from the measured GW signals, one requires template waveforms 
that have to be compared with the observational data using a Bayesian framework~\cite{Veitch:2014wba}
to estimate the intrinsic binary properties such as masses, spins, or deformability, and 
extrinsic parameters such as the sky location or distance. 
Various techniques have been applied to construct such template waveforms, 
including  Post-Newtonian theory (Ref.\ \cite{Blanchet:2013haa} and references therein), 
the effective-one-body framework, e.g.,~\cite{Buonanno:1998gg,Damour:2009wj}, 
numerical-relativity simulations, e.g.,~\cite{Mroue:2013xna,Dietrich:2018phi,Kiuchi:2019kzt}, 
or simply phenomenological descriptions~\cite{Ajith:2007qp,Hannam:2013oca,Dietrich:2017aum,Kawaguchi:2018gvj}. 

For the case of BHNSs, only a limited number of waveform models exist, namely, 
the \texttt{LEA}~\cite{Lackey:2013axa} model and its upgraded \texttt{LEA+} 
version\footnote{Both models have a limited coverage of the parameter space 
with mass ratios between 2 and 5}, the \texttt{PhenomNSBH} model~\cite{Thompson:2020nei}, 
and the \texttt{SEOBNRv4\_ROM\_NRTidalv2\_NSBH} model~\cite{Matas:2020wab}. 
In addition, also more generic effective-one-body models such as \cite{Bernuzzi:2014owa,Hinderer:2016eia,Steinhoff:2016rfi,Nagar:2018zoe,Steinhoff:2021dsn} can describe BHNS systems, but generally, miss a clear BHNS-specific merger morphology of the GW amplitude.
In fact, there is currently no GW model that combines both, the description of tidal effects and BHNS-specific amplitude corrections, with the description of higher-modes which, for the case of BHNS mergers, might be of particular interest to e.g., measure the Hubble parameter~\cite{Vitale:2018wlg,Feeney:2020kxk}.

Given that the upcoming observing runs of Advanced LIGO and Advanced Virgo
with their improved sensitivities will detect numerous compact binary systems, including BHNSs, 
and the limitations considering our capability to model accurately BHNS systems, 
there is a strong interest in further improving GW models. 
Such upgrades require an accurate understanding of the merger process 
and a correct description of possible disruption of the NSs 
within the gravitational field of the BHs, hence, first principle 
numerical-relativity simulations have to be performed. 
To date, the number of existing BHNS simulations is limited and only a few of 
these are accurate enough to be directly used for the calibration of GW models, 
cf.~discussions in~\cite{Thompson:2020nei,Matas:2020wab}.

One of the main difficulties in performing BHNS simulations is the 
construction of proper initial data (ID). However, several efforts have been made 
towards constructing IDs for mixed binaries, e.g., the~\texttt{Spells} code~\cite{Pfeiffer:2002wt,Foucart:2008qt,Tacik:2016zal}
used by the SXS collaboration, the \texttt{TwoPunctures} code and its adaptation towards BHNSs~\cite{Ansorg:2004ds,Clark:2016ppe,Khamesra:2021duu}, the publicly available
\lorene~code~\cite{LoreneCode}, a private version of \lorene~\cite{Kyutoku:2014yba}, and the recently released~\texttt{FUKA}~code~\cite{FukaCode,Papenfort:2021hod}\footnote{This code became public while we were at the end of finishing this article.} based on the \texttt{Kadath} library~\cite{Grandclement:2009ju}.

To follow this line of research and to improve the situation 
with respect to the limited number of BHNS simulations, we perform a set of
new numerical-relativity simulations for BHNS systems 
using the \bam\ code~\cite{Brugmann:2008zz,Thierfelder:2011yi,Dietrich:2015iva,Bernuzzi:2016pie}. 
In the past, \bam\ has shown its capability to perform accurate BNS~\cite{Bernuzzi:2016pie,Dietrich:2018upm,Dietrich:2018phi} and BBH~\cite{Husa:2007hp,Hannam:2010ec,Husa:2015iqa}
simulations, but no BHNS simulations have been performed yet. Here
we demonstrate that \bam\ is capable of performing BHNS simulations
of good quality.

The paper is structured as follows, 
in Sec.~\ref{sec:gr_eqns}, we review the most important equations for our general-relativistic hydrodynamics simulations.
In Sec.~\ref{sec:simu} we discuss \bam's code structure and also the changes required for our BHNS simulations. 
Sec.~\ref{sec:validation} provides test cases to validate our approach
and we compare our results with existing, publicly available BHNS 
simulations performed by the SXS collaboration~\cite{SXS:catalog,Hinderer:2016eia}\footnote{In these works, 
excision initial data is evolved using excision methods for the evolution.}, 
and with simulations performed with SACRA code~\cite{Kyutoku:2010zd}\footnote{In Ref.~\cite{Kyutoku:2010zd} 
puncture initial data is evolved using moving puncture gauge.}.
These comparisons are not only essential to validate our results, 
but also provide one of the first code-comparison studies for BHNSs. 
Finally, Sec.~\ref{sec:summary} summarizes our findings.

Throughout the paper, geometric units are used such that $G=c=1$, 
and in addition, we set $M_\odot=1$.

\section{Equations}
\label{sec:gr_eqns}

Given that we will perform among the first BHNS simulations with 
the~\bam~code~\cite{Brugmann:2008zz,Thierfelder:2011yi,Dietrich:2015iva,
Bernuzzi:2016pie}, we want to review in the following the most important evolution equations. 
We start by assuming the usual 3+1 decomposition of spacetime 
\begin{equation}
\label{eqn:lineelem}
ds ^2 = -(\alpha ^2 - \beta _i \beta ^i) \ dt ^2 + 2 \beta _i
\ dt \ dx ^i + \gamma _{ij} \ dx ^i \ dx ^j .
\end{equation}
Here $\alpha$ and $\beta^i$ are the lapse and shift vector, and
$\gamma _{ij}$ denote the spatial 3-metric. The Einstein equations are
formulated according to the \z4c
formalism~\cite{Bernuzzi:2009ex,Hilditch:2012fp} and
summarized in Sec.~\ref{sec:metric}. General relativistic hydrodynamics
(GRHD) equations are given in
Sec.~\ref{sec:matter} following the flux-conservative formulation
of~\cite{Banyuls:1997zz,Thierfelder:2011yi}.

\subsection{Metric}
\label{sec:metric}
In the \texttt{Z4} formulation, the Einstein equations are
rewritten as
\begin{align}
\label{eqn:z4form}
R _{\alpha \beta} + \nabla _\alpha Z _\beta &+ \nabla _\beta Z _\alpha =
8 \pi \left(T _{\alpha \beta} - \half g_{\alpha \beta} T \right) \\ &+ \kappa _1
\left( t _\alpha Z _\beta + t _\beta Z _\alpha  - (1+\kappa _2) g _{\alpha \beta} \right) t _\gamma Z ^\gamma \non
\end{align}
where $Z _\alpha$ is a four-vector consisting of constraints, $t _\alpha$ is a timelike vector,
and $\kappa _1$, $\kappa _2$ are (constraint) damping parameters. When the constraints vanish, Eq.~(\ref{eqn:z4form})
is equivalent to the standard form of the covariant Einstein equations. 
By introducing a conformal decomposition,
\begin{align}
\label{eqn:z4c1}
\tilde{\gamma}_{ij} &= \chi \gamma _{ij} \ , \\
\tilde{A} _{ij} &= \chi \left( K _{ij} - \frac{1}{3} \gamma_{ij} K \right) \ , \\
\hat{K} &= \gamma ^{ij} K _{ij} - 2 \Theta \ ,
\end{align}
where $\Theta = - n _\alpha Z ^\alpha$, the \z4c~evolution equations are written as
\begin{align}
\label{eqn:z4c2}
\partial _t \chi &= \frac{2}{3} \chi \left(\alpha \left( \hat{K} + 2 \Theta \right) - D _i \beta ^i \right) \\
\partial _t \tilde{\gamma} _{ij} &= -2 \alpha \tilde{A} _{ij} + \beta ^k \partial _k \tilde{\gamma} _{ij} \non \\
& + 2 \tilde{\gamma} _{k(i}\partial _{j)} \beta ^k - \frac{2}{3} \tilde{\gamma} _{ij} \partial _k \beta ^k,
\end{align}
for the metric components,
\begin{align}
\label{eqn:z4c2}
\partial _t \hat{K} &= - D ^i D _i \alpha + \alpha
\left( \tilde{A} _{ij} \tilde{A} ^{ij} + \frac{1}{3} \left( \hat{K} + 2 \Theta \right) ^2 \right) \non \\
&+ 4 \pi \alpha \left( S + E \right) + \beta ^k \partial _k \hat{K} + \alpha \kappa _1 \left( 1 - \kappa _2 \right) \Theta, \\
\partial _t \tilde{A} _{ij} &= \chi \left(- D _i D _j \alpha + \alpha \left( ^{(3)}R _{ij} -8 \pi S _{ij} \right) \right)^{\rm TF} \non \\
&+ \alpha \left( \left( \hat{K} + 2\Theta \right) \tilde{A} _{ij} - 2 {\tilde{A} ^k} _i \tilde{A} _{kj}  \right) + \beta ^k \partial _k \tilde{A} _{ij} \non \\
&+ 2 \tilde{A} _{k(i} \partial _{j)} \beta ^k - \frac{2}{3} \tilde{A} _{ij} \partial _k \beta ^k,
\end{align}
 for the extrinsic curvature components, and
\begin{align}
\label{eqn:z4c3}
\partial _t \tilde{\Gamma} ^i &= -2 \tilde{A} ^{ik} \partial _k \alpha + 2\alpha
\left({\tilde{\Gamma} ^i} _{kl} \tilde{A} ^{kl} - \frac{3}{2} \tilde{A} ^{ik} \partial _k \ln (\chi) \right. \non \\
&\left. - \frac{1}{3} \tilde{\gamma} ^{ik} \partial _k \left(2 \hat{K} + \Theta \right) -8\pi \tilde{\gamma} ^{ik} S _k  \right) + \tilde{\gamma} ^{kl} \partial _k \partial _l \beta ^i \non \\
&+ \frac{1}{3} \tilde{\gamma} ^{ik}  \partial _l \partial _k \beta ^l - 2 \alpha \kappa _1 \left(\tilde{\Gamma} ^i - \bar{\Gamma} ^i \right) + \beta ^k \partial _k \tilde{\Gamma} ^i \non \\
&- \bar{\Gamma} ^k \partial _k \beta ^i + \frac{2}{3} \bar{\Gamma} ^i \partial _k \beta ^k, \\
\partial _t \Theta &= \frac{\alpha}{2} \left( ^{(3)}R - \tilde{A} _{ij} \tilde{A} ^{ij} + \frac{2}{3} \left( \hat{K}
+ 2\Theta \right)^2 \right) \non \\
&- \alpha \left( 8 \pi E + \kappa _1 (2+\kappa _2) \Theta \right) + \beta ^i \partial _i \Theta,
\end{align}
for the remaining variables\footnote{A missing factor of 2 from Eq.~(10) in the `$\kappa _1$'-term
of~\bam~implementation was fixed in the simulations performed in this article. The correction had the effect of reducing
the constraint violations.} with $\tilde{\Gamma} ^i = 2 \tilde{\gamma} ^{ik} Z _k
+ \tilde{\gamma} ^{ij} \tilde{\gamma} ^{kl} \tilde{\gamma} _{jk,l}$
and $\bar{\Gamma} ^i = \tilde{\gamma} ^{kl} {\tilde{\Gamma} ^i} _{,kl}$~cf.~\cite{Hilditch:2012fp}.
The important advantages of the \z4c system are the constraint damping property and that there are no zero-speed characteristic variables in the constraint subsystem. These properties make the \z4c formulation the preferred choice for our numerical simulations presented in this article.

The gauge is specified by the (1+log) lapse~\cite{Bona:1994b} and Gamma-driver-shift 
conditions~\cite{Alcubierre:2002kk,vanMeter:2006vi}:
\begin{align}
\left( \partial _t - \beta ^j \partial _j \right) \alpha &= -\alpha ^2 \mu _L \hat{K}, \\
\left( \partial _t - \beta ^j \partial _j \right) \beta ^i &= \alpha ^2 \mu _S \tilde{\Gamma}^i - \eta \beta ^i.
\end{align}
We set the initial value of the gauge variables to $\alpha = 1$ and $\beta ^i = 0$.
The gauge parameters in our simulations are fixed to $\mu _L = 2/\alpha$,
$\mu _S = 1/\alpha ^2$ and $\eta = 2/M_\text{ADM}$, unless otherwise stated.

\subsection{Matter}
\label{sec:matter}
The GRHD equations are written in the first-order flux-conservative hyperbolic system as
\begin{align}
\label{eqn:grhd1}
\partial _t \vec{q} + \partial _i \vec{f} ^{(i)} (\vec{q}) = \vec{s} (\vec{q})
\end{align}
where the conservative variables, the flux and the source terms are defined as
\begin{align}
\label{eqn:grhd2}
\vec{q} &= \vec{q}(\vec{w}) \equiv \sqrt{\gamma} \left(D, S _k, \tau \right) \ , \non \\
\vec{f} ^i &= \vec{f} ^i(\vec{w}) \equiv \left\lbrace D \left( v ^i
- \frac{\beta ^i}{\alpha} \right), S _j \left( v ^i - \frac{\beta ^i}
{\alpha} \right) + p \delta ^i _j, \right. \non \\
&\left. \tau \left( v ^i - \frac{\beta ^i}
{\alpha} \right) + p v ^i \right\rbrace , \non \\
\vec{s} &= \vec{s}(\vec{w}) \equiv \left\lbrace 0, T ^{\mu \nu}
\left( \partial_\mu g _{\nu j} - {\Gamma ^\sigma} _{\nu \mu} g _{\sigma j}
\right), \right. \non \\
&\left. \alpha \left( T ^{\mu 0} \partial _\mu (\ln\alpha) - T ^{\mu \nu}
{\Gamma ^0} _{\nu \mu}   \right) \right\rbrace.
\end{align}
respectively, and where the conservative variables are defined in terms of the
primitive variables $\vec{w}= \{p,\rho,\epsilon,v ^i\}$:
\begin{align}
\label{eqn:grhd3}
D &\equiv W \rho , \non \\
S _k &\equiv W ^2 \rho h v _k , \\
\tau &\equiv \left( W ^2 \rho h - p \right) -D  . \non
\end{align}
These variables represent the rest-mass density $(D)$, the momentum density
$(S _k)$ and the internal energy  density $(\tau = \rho _\text{ADM} - D)$ as viewed
by Eulerian observers. $v ^i$ is the fluid velocity measured
by the Eulerian observer with
\begin{align}
v ^i = \frac{u ^i}{W} + \frac{\beta ^i}{\alpha} = \frac{1}{\alpha} \left( \frac{u ^i}{u ^0} + \beta ^i\right),
\end{align}
$W$ is the Lorentz factor between the fluid frame and the Eulerian observer,
$W = 1/\sqrt{1 - v ^2} $, with $v ^2 = \gamma _{ij} v ^i v^j$.
The system in Eq.~(\ref{eqn:grhd1}) is closed by an equation of state (EOS) of the
form $p=p(\rho,\epsilon)$. A simple EOS is the $\Gamma-$law $p(\rho,\epsilon) =  (\Gamma-1)\rho \epsilon$,
or its barotropic version $p(\rho) = \kappa \rho ^{\Gamma}$ (polytropic EOS).
Several barotropic zero-temperature NS EOSs
can be fit to acceptable accuracy by piecewise polytropes so that they can be efficiently used in simulations.
In this article we employ two, four, and nine segment fitting piecewise-polytropic
models~\cite{Read:2009yp} for four of the five EOSs used\footnote{For the two EOSs (EOS1 \& EOS3)
we fit the entire table publicly available in~\cite{Annala:2019puf}
and not use the crust model as in~\cite{Read:2009yp}. }.
Additionally, we add a thermal pressure component, $p_{\rm th} = \rho \epsilon (\Gamma_\text{th}-1)$ 
with $\Gamma_{\rm th}=1.75$, to the cold pressure \cite{Zwerger:1997}.
The system in Eq.~(\ref{eqn:grhd1}) is strongly hyperbolic provided that the EOS is causal, i.e., the sound speed is 
less than the speed of light.\\
More specifically, we use the following equations of state:
\bi
\item EOS1: a sub-conformal EOS with a crossover transition, leading to sizable quark matter
(QM) cores in massive NSs (R$=$6.4 km for M$_{\rm max}$$\sim 1.99 \; \Msun$)~~\cite{Annala:2019puf}.
\item EOS3: a high-$c_s$ EOS with a strong first-order phase transition, leading to no QM cores~~\cite{Annala:2019puf},
\item SLy: derived via the Skyrme-type effective nuclear interaction SLy  \cite{Douchin:2001sv},
\item Poly$\kappa 101.45\Gamma 2$: a single polytrope with $\kappa = 101.45$ and $\Gamma = 2$,
\item HB: a piecewise polytrope consisting of a simple crust connected to a single piece 
for the core with $\Gamma=3$, e.g.~\cite{Kyutoku:2010zd}. 
\ei

\section{Numerical Methods}
\label{sec:simu}

\subsection{The \bam~code}

\begin{figure}[htb]
\centering
\includegraphics[width=0.5\textwidth]{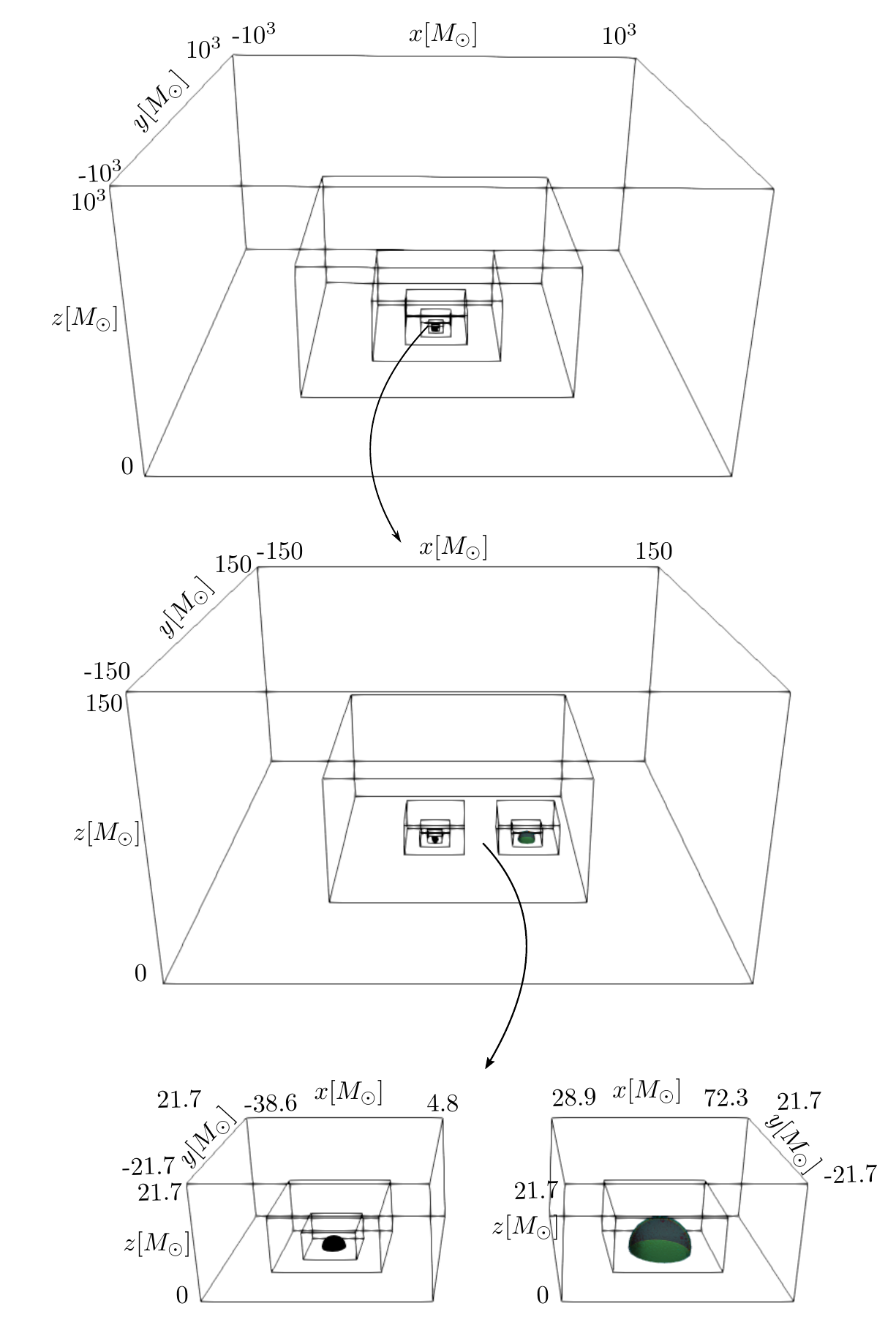}
\caption{A schematic overview of the grid setup employed in the \bam~code.
For BHNS systems the BH is resolved with extra refinement levels
as compared to the NS. The simulations presented in this article have
one additional refinement level for the BH. The finest levels show the BH horizon
as `black' contour and the NS as the `green' contour. In all simulations
we employ bitant symmetry to save computational costs.}
\label{fig:bam_amr}
\end{figure}

\begin{table}[tp]
\caption{Grid configurations. The first column gives the configuration name in the form `EOSQmassratio'
where mass ratio is defined as Q$ = M ^{BH}/M ^{NS} _g$ and the `$\uparrow$' indicates a (aligned)
spinning BH. The next eight columns give the number of levels $L$, the number of moving box levels $L_{\rm mv}$,
the number of points in the nonmoving boxes $n$, the number of points 
in the moving boxes $n_{\rm mv}$, the grid spacing $h_{9}$ ($h_{5}$) in the finest level
covering the NS, the grid spacing $h_{10}$ ($h_{6}$) in the finest level
covering the BH, the grid spacing $h_0$ in the coarsest level,
and the outer boundary position $R_0$. The grid spacing and the
outer boundary position are given in units of $M_{\odot}$.}
\label{tab:grid} 
\centering
\begin{footnotesize}
\begin{tabular}{ccccccccc}
\toprule
Name & $L$ & $L_{\rm mv}$ & $n$ & $n_{\rm mv}$ & $h_{9}$ & $h_{10}$ & $h_0$ & $R_0$ \\ \hline
EOS1Q2.95 & 11 & 3 & 256 & 128 & 0.156 & 0.078 & 80. & 10240. \\
EOS3Q2.98 & 11 & 3 & 256 & 128 & 0.156 & 0.078 & 80. & 10240. \\
SLyQ2 & 11 & 3 & 256 & 128 & 0.156 & 0.078 & 80. & 10240. \\
SLyQ2$^{\uparrow}$ & 11 & 3 & 256 & 128 & 0.156 & 0.078 & 80. & 10240. \\
SLyQ2.84 & 11 & 3 & 256 & 128 & 0.156 & 0.078 & 80. & 10240. \\
SLyQ2.84$^{\uparrow}$ & 11 & 3 & 256 & 128 & 0.156 & 0.078 & 80. & 10240. \\
Poly$\kappa 101.45\Gamma 2$Q2 & 11 & 3 & 256 & 128 & 0.188 & 0.094 & 96. & 12288. \\
HBQ2-R1 & 11 & 3 & 192 & 96 & 0.208 & 0.104 & 106.67 & 10240. \\
HBQ2-R2 & 11 & 3 & 256 & 128 & 0.156 & 0.078 & 80. & 10240. \\
HBQ2-R3 & 11 & 3 & 288 & 144 & 0.139 & 0.069 & 71.11 & 10240. \\
\toprule
Name & $L$ & $L_{\rm mv}$ & $n$ & $n_{\rm mv}$ & $h_{5}$ & $h_{6}$ & $h_0$ & $R_0$ \\ \hline
SLyQ4.76\footnote{Setup used for tests.} & 7 & 4 & 160 & 128 & 0.125 & 0.063 & 4. & 320. \\ \hline \hline
\end{tabular}
\end{footnotesize}
\end{table}

The computational domain is divided into a hierarchy of cell 
centered nested Cartesian grids with refinement factor of $2$.
The hierarchy consists of $L$ levels of refinement indexed by $l = 0,...,L-1$.
Each level has one or more Cartesian grids with constant grid spacing
$h_l$ and $n$ (or $n^{\rm mv}$) points per direction.
The grids are properly nested such that the coordinate extent
of any grid at level $l,l >0$, is completely covered by the
grids at level $l-1$. Refinement levels $l > l^{\rm mv}$
can be dynamically moved and follow the motion of the compact 
objects according to ``moving boxes'' technique~\cite{Brugmann:2008zz}. 
In this article, we set $l^{\rm mv} = 7~(2)$. Furthermore, to adequately resolve the BH and the region around it, extra refinement levels can be added only for the BH. In all the simulations presented in this article we add one extra level for the BH. Figure~\ref{fig:bam_amr} shows a schematic of the refinement grid structure for a typical BHNS simulation in \bam. Moreover, we use bitant symmetry, i.e., reflection across z=0 plane, 
in the simulations to half the computational costs. 
In Tab.~\ref{tab:grid} we list the grid configurations 
used in this article.

\begin{figure}[htb]
\centering
\includegraphics[width=\linewidth]{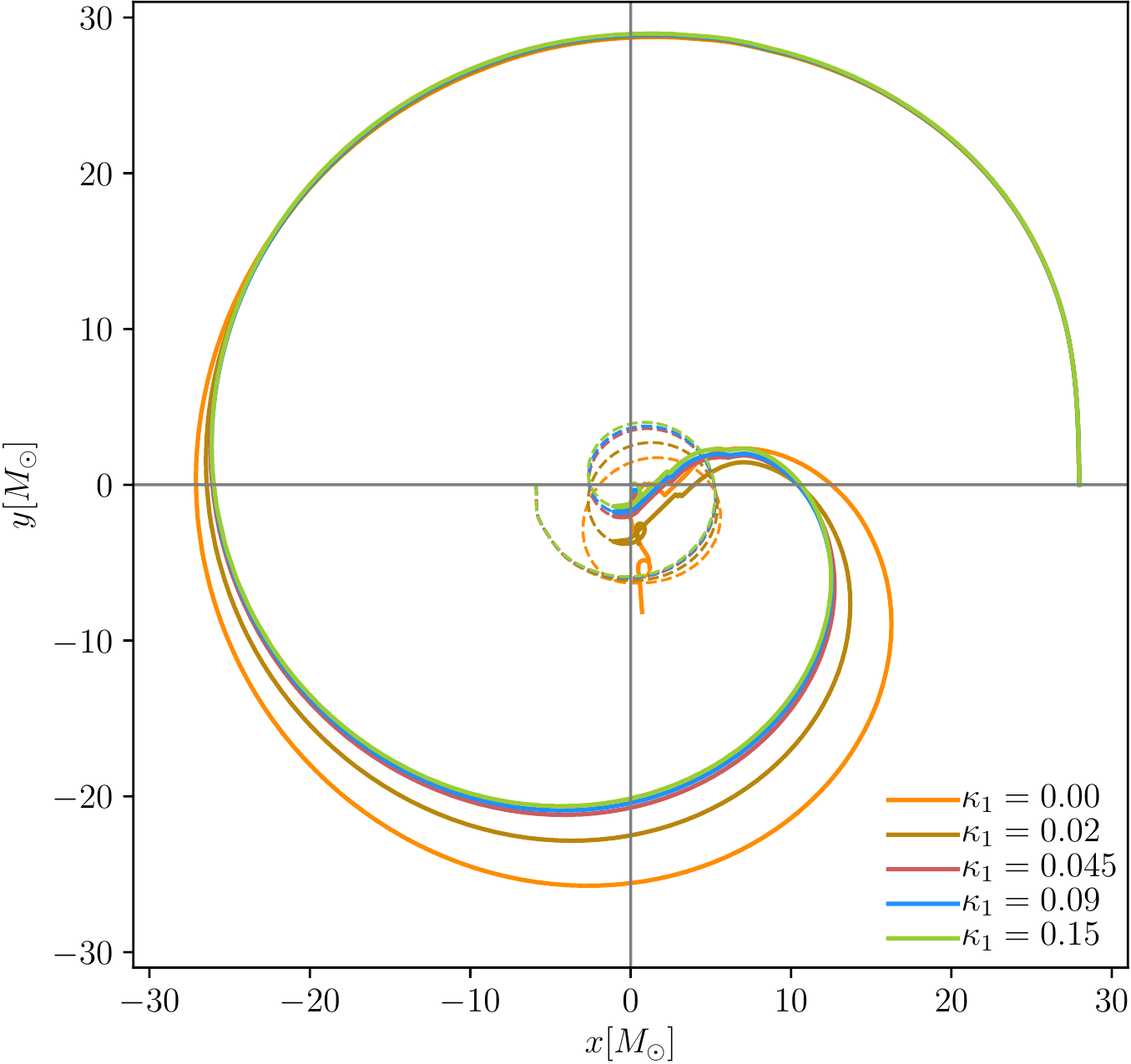}
\caption{Puncture tracks of the BH (dashed line) and the NS (solid line) for the tests (setup SLyQ4.76)
with different constraint damping parameters using the \z4c formulation.
Larger values for the $\kappa _1$ parameter help in reducing the drift of the center-of-mass
that is introduced due to initial constraint violations inside the BH.}
\label{fig:tests}
\end{figure}

The IDs are evolved with the \z4c formulation of the Einstein equations
for the evolution system as described in Sec.~(\ref{sec:metric}).
Constraint damping scheme with values of $\kappa _1 \in [0.045-0.065]$\footnote{We also
set $\kappa _1 = \{0.0,0.02,0.09,0.15\}$ for some of the tests.}
and $\kappa _2 = 0$ are used. These values are used based on the suggestions in the
detailed 1D numerical analysis of Ref.~\cite{Weyhausen:2011cg} and the tests performed in this article, cf.~Fig.~\ref{fig:tests}.
The combined use of artificial dissipation and constraint damping terms is important (and in some cases essential) to avoid instabilities arising from constraint violating ID inside the BH.~\bam~ implements a Kreiss-Oliger dissipation of the form 0.5 $\times$ 2$^{-6}$ ($\Delta x $)$^6$ ($\partial _x ^6 + \partial _y ^6 + \partial _z ^6$) for all the gravitational field variables at each intermediate Runge–Kutta timestep, where $\Delta x $ is the grid separation~\cite{Gustafsson:1995}.

We use Sommerfeld boundary conditions~\cite{Hilditch:2012fp}\footnote{We note that using radiative
boundary conditions for the \z4c system on box-boundaries is problematic as it results in non-convergent
reflections that travel towards the system during the evolution. This is also the reason for using larger
outer boundaries in our simulations. We also tried the scheme described in Sec.~IIIA of~\cite{Kyutoku:2014yba}
to damp the reflections but it did not work for our simulations, possibly, because of some subtle differences
in \bam~and \texttt{SACRA}.}, the method-of-lines
for the time integration with fourth-order Runge-Kutta integrator and fourth-order
finite differences for approximating spatial derivatives. A Courant-Friedrich-Lewy (CFL) factor of $0.25$
is employed for all runs~\cite{Brugmann:2008zz,Cao:2008wn}. Moreover, the time stepping utilizes the Berger-Collela scheme, enforcing mass conservation across the refinement boundaries~\cite{Berger:1984zza,Dietrich:2015iva}.
GWs are extracted using the curvature scalar $\Psi _4$, cf.\ Sec.~III of~\cite{Brugmann:2008zz}.

The numerical fluxes for the GRHD system,
as described in Sec.~(\ref{sec:matter}), are constructed with a flux-splitting approach based on the local Lax-Friedrich (LLF) scheme. 
We perform the flux reconstruction with 
a fifth-order WENOZ algorithm~\cite{Borges:2008a} on 
the characteristic fields~\cite{Jiang:1996,Suresh:1997,Mignone:2010br} to obtain high-order convergence~\cite{Bernuzzi:2016pie}.
For low density regions and around the moment of merger,
we switch to a primitive reconstruction scheme that is more stable but less accurate, i.e., 
from a scheme with potentially higher order convergence that uses the characteristic fields 
to a second-order LLF scheme that simply uses, the primitive variables~\cite{Bernuzzi:2016pie}.

In our simulations the NS is surrounded by an artificial atmosphere, 
e.g.,~\cite{Thierfelder:2011yi,Font:1998hf,Dimmelmeier:2002bk}.
The artificial atmosphere outside of the star is chosen
as a fraction of the initial central density of the star
as $\rho_{atm} \equiv f_{atm} \cdot \rho_c(t = 0)$. The atmosphere
pressure and internal energy is computed by employing
the zero-temperature part of the EOS. The fluid velocity
within the atmosphere is set to zero. At the start
of the simulation, the atmosphere is added before the
first evolution step. During the recovery of the primitive
variables from the conservative variables, a point
is set to atmosphere if the density is below the threshold
$\rho_{thr} \equiv f_{thr} \cdot \rho_{atm}$.
In this article, we are using $f_{atm} = 10^{-13}$ and $f_{thr} = 10^2$
in all the configurations.

\subsection{Upgrades to simulate BHNS systems}

We construct BHNS IDs using the public version of the~\lorene~code. \lorene\ employs multi-domain
spectral methods to obtain the solution to the elliptic equations~\cite{Bonazzola:1998qx,Grandclement:2001ed}.
The first BHNS IDs constructed using \lorene~are described in~\cite{Grandclement:2006ht}.
Due to the modular architecture of \bam~and~\lorene, both codes have been easily extended to
read and interpolate the spectral ID onto the Cartesian grid of \bam.
To import the spectral configurations from~\lorene~onto our Cartesian simulation grid,
we first construct our simulation grid and note the positions of each grid point. Then we evaluate the geometric and the hydrodynamic fields at these positions based on their spectral coefficients. Lastly, the excised BH region is filled with constraint-violating ID, using the ``smooth junk'' technique~\cite{Etienne:2007hr}.\\

As \lorene~uses the excision technique for BHs when constructing IDs, the BH interior is removed from the computational
domain to avoid pathologies due to the physical singularity and one applies appropriate
inner boundary conditions at the excision surface~\cite{Gourgoulhon:2001ec,Grandclement:2001ed}.
Within \bam{}, however, we are using the moving puncture approach so that valid data are also required in the excised region of the ID. 
To circumvent this issue we fill the excised interior with arbitrary but smooth data and evolve it with standard puncture gauge choices as described in Sec.~(\ref{sec:metric}).
During the evolution, we then use the constraint damping properties of the \z4c~formulation to reduce effects of the initial constraint violation inside the BH~\cite{Gundlach:2005eh,Weyhausen:2011cg,Cao:2011fu}.

For filling the excised interior, we perform a seventh-order polynomial
extrapolation of all the field values radially
using uniform points from $r \geq r_{AH}$. The extrapolating polynomial
is given explicitly by Lagrange's formula,
\begin{equation}
f _{n-1} (r _i) = \sum ^n _{i=1} L ^n _i (r) f(r _i)
\end{equation}
where $L ^n _i$ is
\begin{equation}
L ^n _i (r) = \prod ^n _{\substack{j = 1 \\ j \neq i}} \frac{r - r _j}{r _i - r _j} . \nonumber
\end{equation}
The Lagrange formula is not implemented straightforwardly, but instead via Neville's algorithm since it is computationally more efficient;
the procedure is described in detail in Ref.~\cite{Press:1992}.

\section{Code Validation}
\label{sec:validation}

\begin{table*}[htb]
\caption{BHNS configurations. The first column refers to the configuration name.
The next five columns provide the physical properties of the BH and the NS:
the BH area mass $M^{BH}$, the BHs'
dimensionless spin magnitude $\chi^{BH}$ (note we only achieved maximal spins of 0.4
in our tests), baryonic mass of the NS $M^{NS}_b$, gravitational
mass of the NS $M^{NS}_g$ and the NS's compactness $\mathcal{C}$. 
The last four columns give the
residual eccentricity $e$ in the ID, the initial GW frequency $M\omega ^0 _{22}$,
the Arnowitt-Deser-Misner (ADM) mass $M_\text{ADM}$, and the ADM angular
momentum $J_\text{ADM}$. `-' marks the unavailability of eccentricity estimate due to a very short evolution.}
\label{tab:config}
\centering
\begin{tabular}{cccccccccc}
\toprule
Name & $M^{BH}$ & $\chi^{BH}$ & $M^{NS}_b$  & $M^{NS}_g$ & $\mathcal{C} (\frac{M}{R})$ & $e$ & $M\omega^0 _{22}$ & $M_\text{ADM}$ & $J_\text{ADM}$ \\ 
\hline 
 SLyQ4.76 & 6.45 & 0 & 1.5 & 1.354 & 0.174 & - & 0.16525 & 7.7306 & 28.53 \\
 EOS1Q2.95  & 5 & 0 & 1.9 & 1.694 & 0.195 & 0.014 & 0.05441 & 6.6413 & 32.46 \\
 EOS3Q2.98 & 5 & 0 & 1.9 & 1.679 & 0.202 & 0.015 & 0.05424 & 6.6268 & 32.19 \\
 SLyQ2   & 2.7 & 0 & 1.5 & 1.354 & 0.174 & 0.009 & 0.05530 & 4.0169 & 14.01\\
 SLyQ2$^{\uparrow}$   & 2.7 & 0.4 & 1.5 & 1.354 & 0.174 & 0.011 & 0.05517 & 4.0692 & 16.86 \\
 SLyQ2.84  & 3.85 & 0 & 1.5 & 1.354 &  0.174 & 0.011 & 0.07754 & 5.1543 & 18.61 \\
 SLyQ2.84$^{\uparrow}$ & 3.85 & 0.4 & 1.5 & 1.354 & 0.174 & 0.018 & 0.07708 & 5.2283 & 24.21\\
 Poly$\kappa 101.45\Gamma 2$Q2 & 2.8 & 0 & 1.509 & 1.403 & 0.145 & 0.006 & 0.03696 & 4.1728 & 16.56 \\
 HBQ2 & 2.7 & 0 & 1.493 & 1.350 & 0.172 & 0.009 & 0.05522 & 4.0129 & 13.97 \\
\hline 
\hline
\end{tabular}
\end{table*}

To test the validity of our evolution of the filled BHNS system,
we perform a comparison with the \texttt{SXS:BHNS:0002} setup from the SXS collaboration's
BHNS catalog~\cite{SXS:catalog} and the setup \texttt{HBQ2M135}
obtained with the \texttt{SACRA} code~\cite{Kyutoku:2010zd}. Apart from those setups we also 
evolve other configurations with varying mass ratio, EOS, and spin of the BH.
All these setups are tabulated in Tab.~\ref{tab:config}.

\subsection{Constraints and Mass Conservation}

\begin{figure}[htb]
\centering
\includegraphics[width=\linewidth]{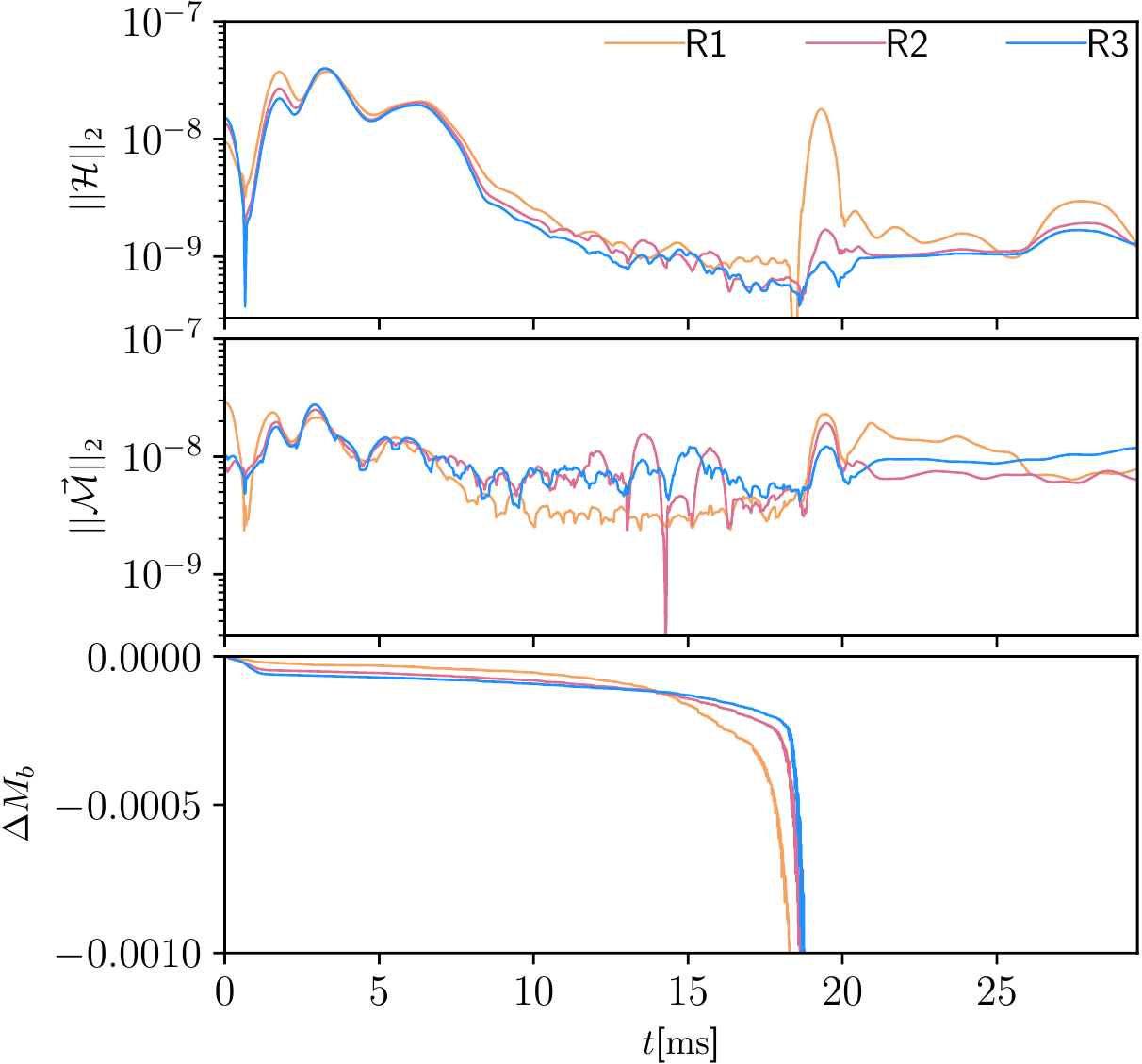}
\caption{The upper panel shows the $L^{2}$ volume norm of the Hamiltonian 
constraint, $||\mathcal{H}||_2$. The middle panel shows the
Euclidean norm of the $L^{2}$ volume norms of the Cartesian components of the
momentum constraint, $||\vec{\mathcal{M}}||_2 = \sqrt{||\mathcal{M}^x||^2 _2 +
||\mathcal{M}^y||^2 _2 + ||\mathcal{M}^z||^2 _2}$. The bottom panel shows
the evolution of the error in the baryonic mass and stays below 0.01\% until
the merger. All the quantities here are
evaluated on refinement level 3 for HBQ2 setup.}
\label{fig:constraints}
\end{figure}

\paragraph*{Einstein Constraints:}
In Fig.~\ref{fig:constraints}~(top and middle panels) we show the evolution of the Hamiltonian and Momentum constraint violations for the HBQ2 setup using different resolutions; cf.~Tab.~\ref{tab:grid} for details.
While we find a monotonic decrease of the Hamiltonian constraint for increasing resolution and that the Hamiltonian constraints decrease over time due to constraint damping, the Momentum constraint violations stay at the level of the initial data without noticeable change during the evolution.

\begin{figure*}[htpb]
\centering
\includegraphics[width=\linewidth]{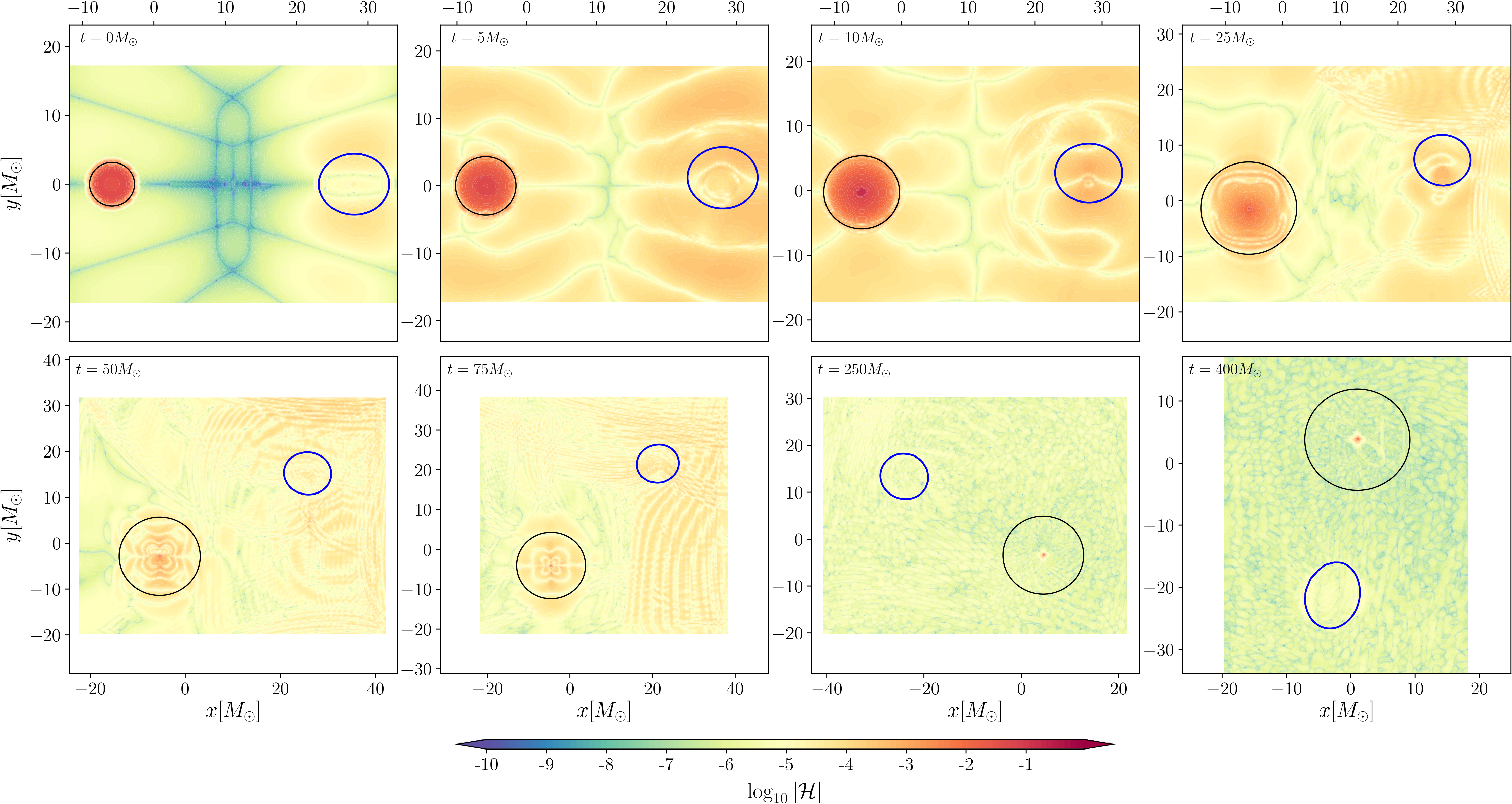}
\caption{Hamiltonian constraint evolution for the SLyQ4.76
configuration. The black contour marks the apparent horizon of the BH and the blue contour marks
the `surface' of the star and corresponds to \mbox{$\rho = 1.462314\times 10^{14}$ g$\cdot$cm$^{-3}$}.
The constraint is computed on the moving level 4 which covers both
the compact objects. The apparent horizon radius changes as it is time dependent due to gauge evolution.}
\label{fig:ham_const2dBHNS}
\end{figure*}

For a more detailed understanding about the impact of the BH stuffing and the transition of the early part during the 
numerical simulation, we present in Fig.~\ref{fig:ham_const2dBHNS} the Hamiltonian constraint around the BHNS system at different 
timesteps for the SLyQ4.76 configuration. As evident from the $t=0$ representation, there are large constraint violations due to the filling of the BH. These constraint violations are within the apparent horizon (marked as black contour).
During the course of the evolution, the constraint violation is decreasing inside the BH due to the transition towards the moving puncture gauge. At latest around $\sim 250M_\odot$, we find that only the puncture shows large constraint violations. 

Considering the evolution of the NS, we find that generally the constraint violation around 
and inside the NS (blue contour) are not noticeably larger than compared to the surrounding spacetime. 
Finally, it is worth pointing out that we notice effects of the grid structure of the initial data 
solver, which is clearly visible in the first two panels, 
and that we see small reflections of the constraint violation, cf.\ panel corresponding to $t=75M_\odot$. 

Overall, while the constraint damping properties of the \z4c evolution scheme lead to a 
reduction of the constraint violations even within the stuffed BH, 
we  do find that small constraint violations leave the inner part of the BH.
Therefore, in contrast to previous works, e.g.~\cite{Etienne:2008re,Etienne:2007hr,Faber:2007dv},
it seems that the exact stuffing and BH filling does have an influence on the dynamical evolution.


\paragraph*{Mass Conservation:}

The bottom panel of Fig.~\ref{fig:constraints}~shows the difference of the baryonic mass during the evolution. Interestingly, we find a small decrease of the baryonic mass during the first few milliseconds of our simulation. This decrease is increasing with resolution and spoils the convergence. After this decrease the mass conservation increases up to the merger of the system when the NS gets disrupted. 
Because of our particular gauge choice and the usage of an artificial atmosphere, mass is not part of the computational domain once it falls inside the BH~\cite{Thierfelder:2010dv,Dietrich:2014cea,Dietrich:2014wja}.

\subsection{Gravitational-Wave Accuracy}

\begin{figure*}[htpb]
\centering
\includegraphics[width=\linewidth]{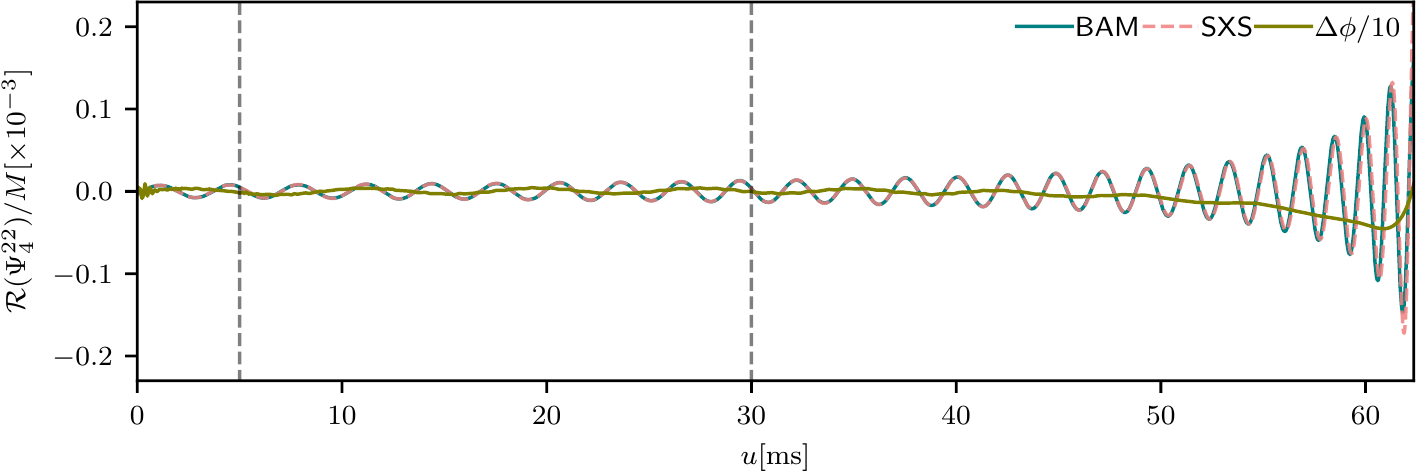}
\caption{Comparison of $\Psi _4 ^{22}$ from SXS catalog (ID-\texttt{SXS:BHNS:0002})
with \bam's Poly$\kappa 101.45\Gamma 2$Q2 setup. 
The alignment interval is marked with vertical dashed lines.}
\label{fig:comparison_SXS}
\end{figure*}

\begin{figure}[htpb]
\includegraphics[width=\linewidth]{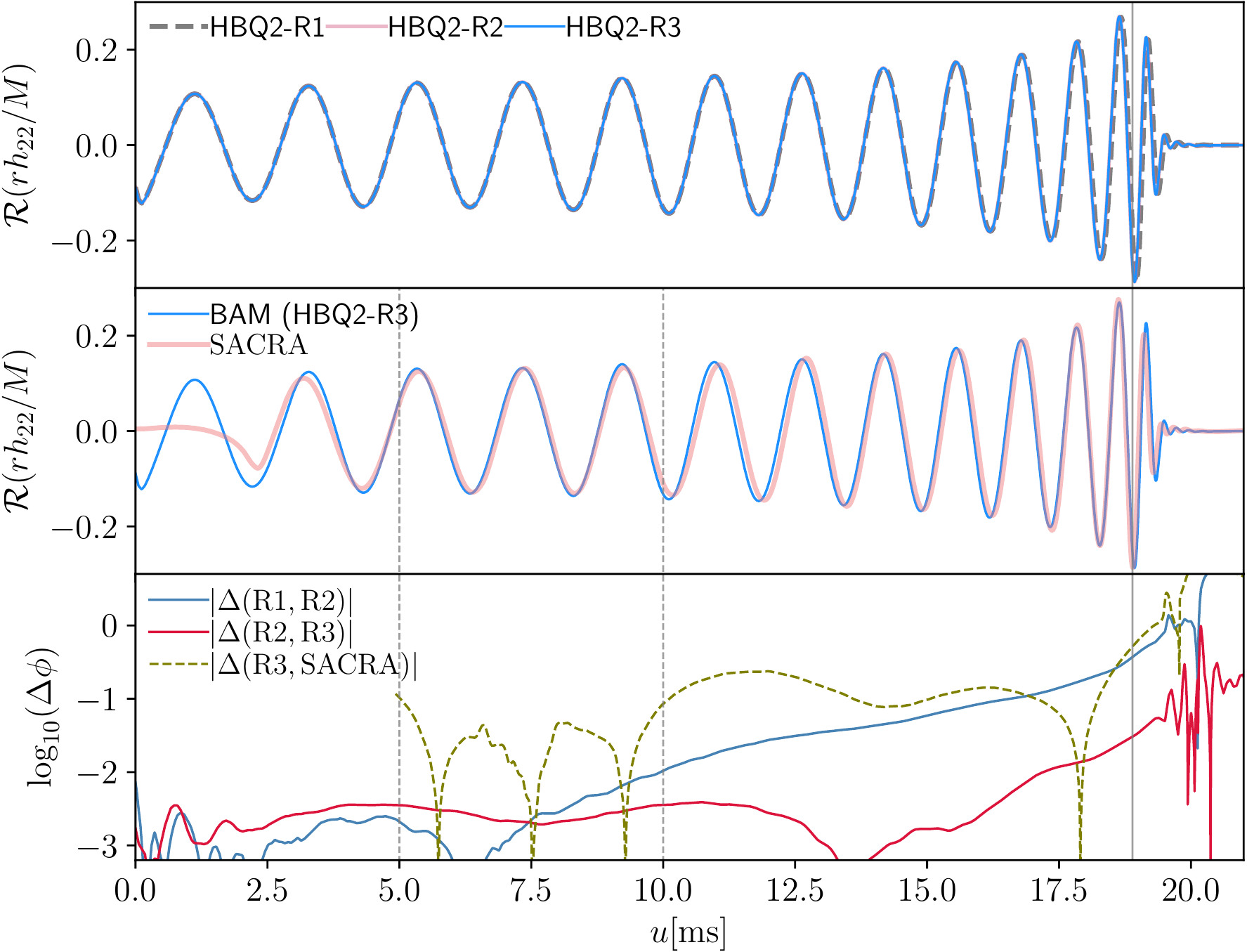}
\caption{\textit{Top:} Waveform ($rh _{22}$) comparison among the various \bam~resolutions
for the HBQ2 configuration. 
\textit{Middle:} Comparison of $rh _{22}$
from \texttt{SACRA} (\texttt{HBQ2M135}~\cite{Kyutoku:2010zd}) with \bam's HBQ2-R3 configuration.
\textit{Bottom:} Phase difference among the various \bam~resolutions for the HBQ2 configuration as well as between \texttt{SACRA} and
\bam's HBQ2-R3 configuration. Overall, the phase difference with the \texttt{SACRA} waveform  and for
the lower \bam~resolution (R1-R2) stays $\lesssim 0.5$ rad until the merger whereas for
the higher \bam~resolution (R2-R3) it stays $\lesssim 0.04$ rad until the merger. The alignment interval is marked with vertical dashed lines and the moment of merger is marked by the solid vertical line.}
\label{fig:comparison_SACRA}
\end{figure}

Figure~\ref{fig:comparison_SXS} shows a comparison of the (2,2)-mode of the curvature scalar $\Psi _4$
for the \bam~evolved Poly$\kappa 101.45\Gamma 2$Q2 setup and the \texttt{SXS:BHNS:0002}
setup from the SXS catalog\footnote{Unfortunately, due to a technical problem the BAM simulation 
could not be continued beyond the moment of merger. To avoid rerunning this long (and computationally expensive) simulation, 
we decided to compare $\Psi_4$ instead of $h$, whose computation would require the entire simulation
including the postmerger part.}.
Comparing the phase difference between our new \bam~simulation and 
\texttt{SXS:BHNS:0002}, we find phase difference up to the end of the simulation 
(which corresponds to the merger) of up to $\sim 0.5$ rad.

Figure~\ref{fig:comparison_SACRA}, top panel, shows the comparison of the GW strain $rh _{22}$ from \bam's HBQ2 simulation
for the three employed resolutions R1, R2, and R3. The middle panel shows the comparison of
$rh _{22}$ from \bam's HBQ2-R3 simulation with the \texttt{HBQ2M135}
setup from \texttt{SACRA}. Here the \texttt{SACRA} waveform has been aligned with
the \bam~waveform in the window marked by vertical dashed-lines as shown in the middle panel
of Fig.~\ref{fig:comparison_SACRA}. The bottom panel shows the phase difference for this comparison (olive dashed-line)
is $\lesssim 0.5$ rad until the merger. It also shows the phase difference among different \bam~resolutions for the HBQ2 configuration.
We note that we do not find a clear convergence order for the GW phase,
but that, in particular, during the last cycles the phase difference between resolutions R2 and R3 is significantly smaller than between R1 and R2.
Furthermore, the overall phase difference is with 
$\approx 0.5\ \rm{rad}$ for the two lowest resolutions 
and $\approx 0.04\ \rm rad$ for the two highest resolutions, surprisingly small. 
Hence, we suggest that there are two possible origins for the missing convergence: 
(i) The stuffing of the BH at $t=0$ adds a constraint violation that to some extend leave the 
apparent horizon and effects the overall convergence properties. 
Such an effect will be investigated through the comparison with another type of initial data 
that either uses a different stuffing formalism or, ideally, 
uses puncture initial data as in Refs.~\cite{Kyutoku:2010zd}. 
(ii) The phase error is overall smaller than during our previous BNS simulations, 
which could indicate that the dominant second order 
error found in previous simulations is absent or suppressed in the case of our BHNS simulations. 
To investigate this options, we would also need further simulations 
that go beyond the computational resources available to us for this project. 

\subsection{Example Simulations}

\begin{figure}[htpb]
\centering
\includegraphics[width=\linewidth]{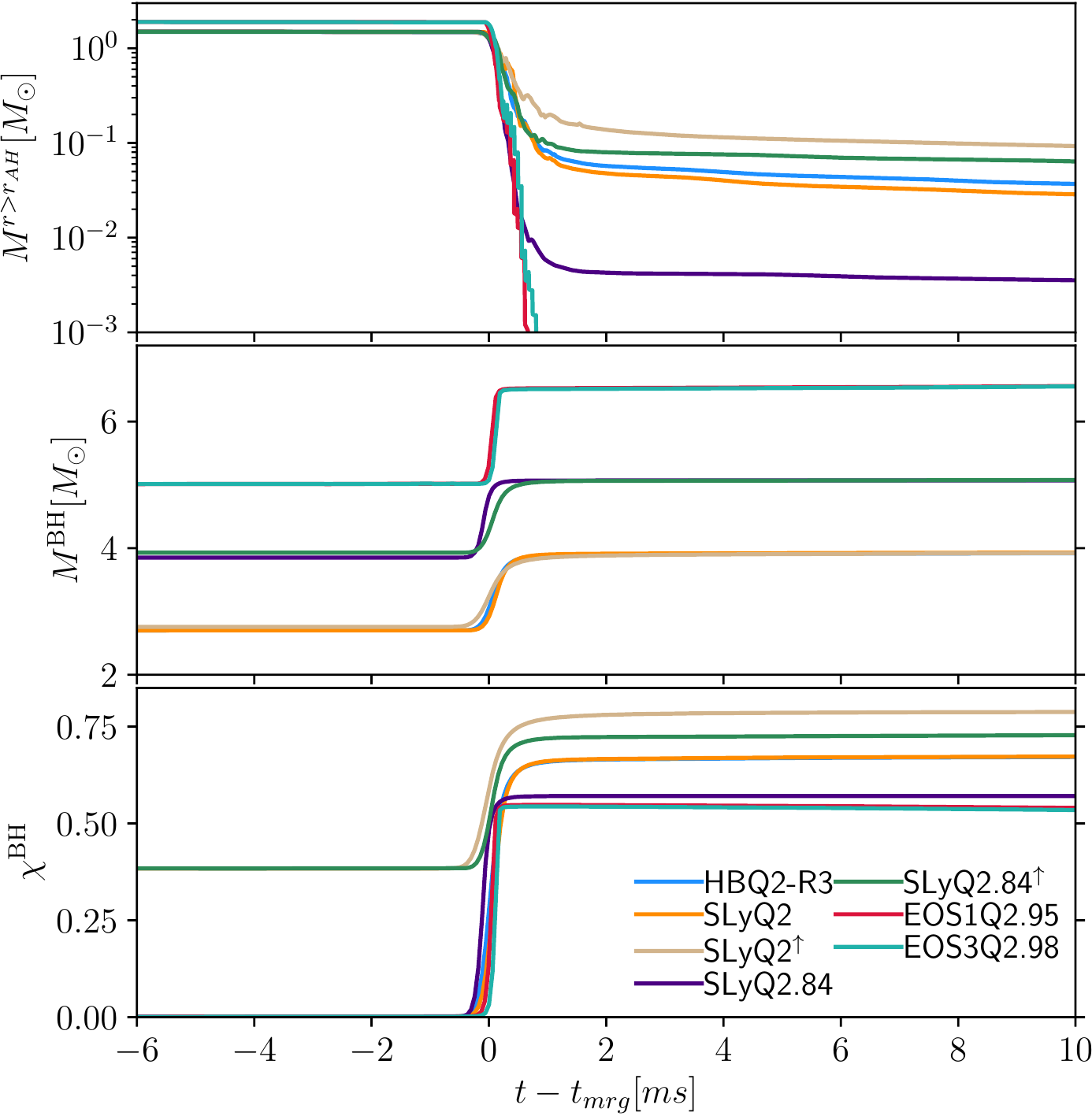}
\caption{Evolution of the rest mass of the material located outside the apparent horizon (\textit{top}) ,
horizon mass (\textit{middle}) and the dimensionless spin of the 
BH (\textit{bottom}) with appropriate time shift; in these plots the time at the 
onset of the merger is taken as the time origin.}
\label{fig:postmerger_matter}
\end{figure}

\begin{figure*}[t]
\begin{subfigure}{\linewidth}
\centering
\includegraphics[width=\linewidth]{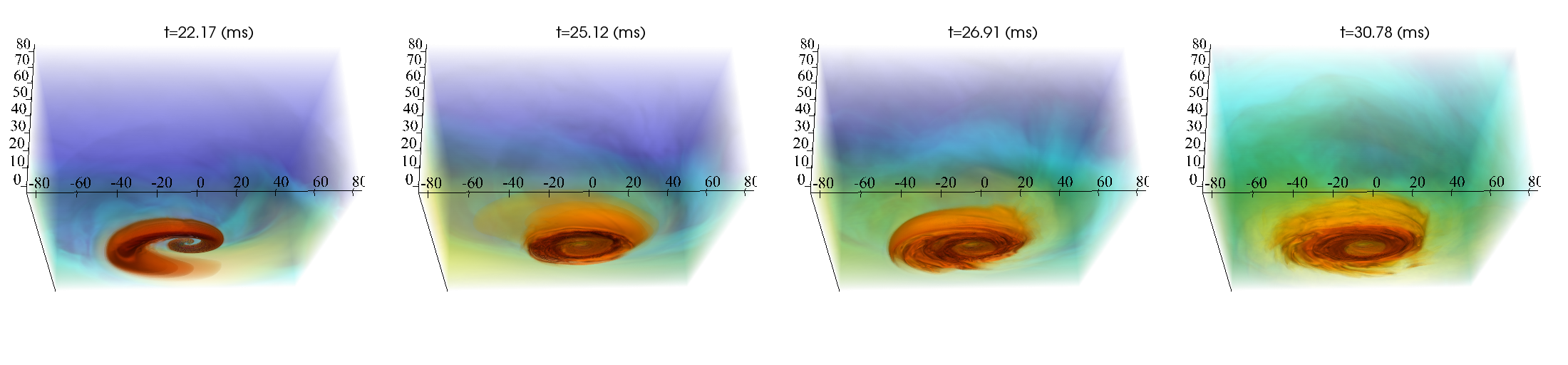}
\end{subfigure}\hfil

\medskip

\begin{subfigure}{\linewidth}
\centering
\includegraphics[width=\linewidth]{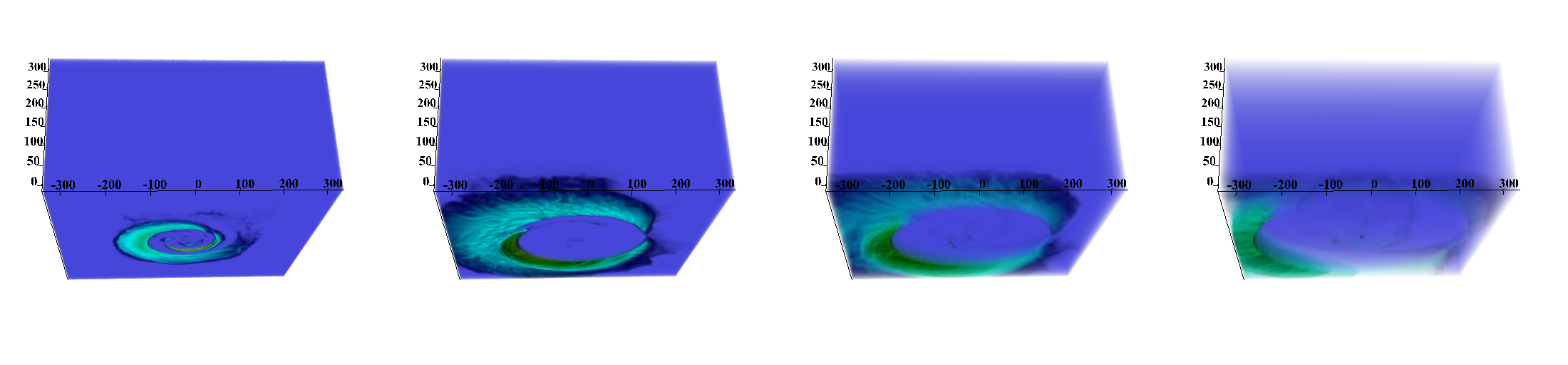}
\end{subfigure}

\medskip

\begin{subfigure}{\linewidth}
\centering
\includegraphics[width=0.6\linewidth]{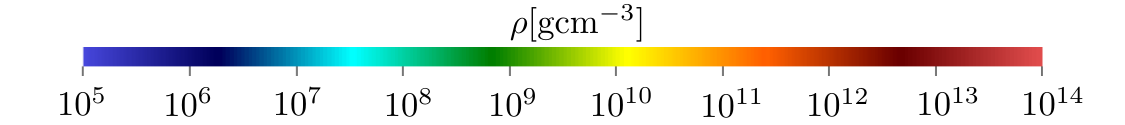}
\end{subfigure}
\caption{Baryon density ($\rho$) evolution of the bound (\textit{Top}) and unbound matter (\textit{Bottom})
for the SLyQ2$^{\uparrow}$ configuration at selected timesteps. The 3D-volume consists only of the half-volume that
we evolve in the simulations. Specifically, top panel domain consists of $x \in [-80 M _\odot, 80 M
_\odot]$, $y \in [-80 M _\odot, 80 M _\odot]$, $z \in [0 M _\odot, 80
M _\odot]$ and bottom panel domain consists of $x \in [-320 M _\odot,
320 M_\odot]$, $y \in [-320 M _\odot, 320 M_\odot]$ and $z \in [0
M_\odot, 320 M_\odot]$. The merger for this setup happens $\sim$ 22 ms.}
\label{fig:3Dmatter}
\end{figure*}

\begin{table}[htpb]
\caption{Ejecta and Merger remnants. The first column gives the configuration name and
the next five columns list the key quantities: for the ejecta; the unbound mass $M_\text{ej}$, its
mass-weighted velocity $\bar{v}_{\rm ej}$; and for the merger remnants, the disk-mass $M^{r>r_\text{AH}}$,
the final BH mass $M^\text{BH}$ and its dimensionless spin $\chi^\text{BH}$. All the quantities are
computed $\sim$ 10 ms after the merger. `-' marks the unavailability of data.}
\label{tab:remnant}
\centering
\begin{tabular}{ccccccc}
\toprule
Name & $M_\text{ej}$ & $\bar{v}_{\rm ej}$ & $M^{r>r_\text{AH}}$ & $M^\text{BH}$ & $\chi^\text{BH}$\\
     & \begin{footnotesize} $[10^{-3}M_{\odot}]$  \end{footnotesize} & \begin{footnotesize} $[c]$  \end{footnotesize} & \begin{footnotesize} $[M_{\odot}]$  \end{footnotesize}&  \begin{footnotesize} $[M_{\odot}]$  \end{footnotesize} & \\
\hline
SLyQ4.76\footnote{Quantities here are reported $\sim$5 ms after the merger where the data was available.}  & $<$0.1 & - & $<$10$^{-5}$ & 8.103 &  0.384 \\
 EOS1Q2.95  & 0.13 & 0.21 & 0.0004 & 6.559 &  0.540 \\
 EOS3Q2.98  & 0.16 & 0.24 & 0.0002 & 6.553 &  0.535 \\
 SLyQ2 & 0.37 & 0.15 & 0.0286 & 3.929 & 0.673  \\
 SLyQ2$^{\uparrow}$  & 2.45 & 0.15 & 0.0927 & 3.923 & 0.788  \\
 SLyQ2.84 & 0.30 & 0.18 & 0.0035 & 5.070 & 0.571  \\
 SLyQ2.84$^{\uparrow}$  & 5.04 & 0.22 & 0.0637 & 5.075 & 0.728  \\
 HBQ2-R1 & 0.95 & - & 0.0427 & 3.919 & 0.672  \\
 HBQ2-R2 & 0.59 & 0.20 & 0.0382 & 3.919 & 0.672  \\
 HBQ2-R3 & 0.44 & 0.14 & 0.0369 & 3.921 & 0.672  \\
\hline 
\hline
\end{tabular} 
\end{table}

\begin{figure*}[htpb]
\centering
\includegraphics[width=\linewidth]{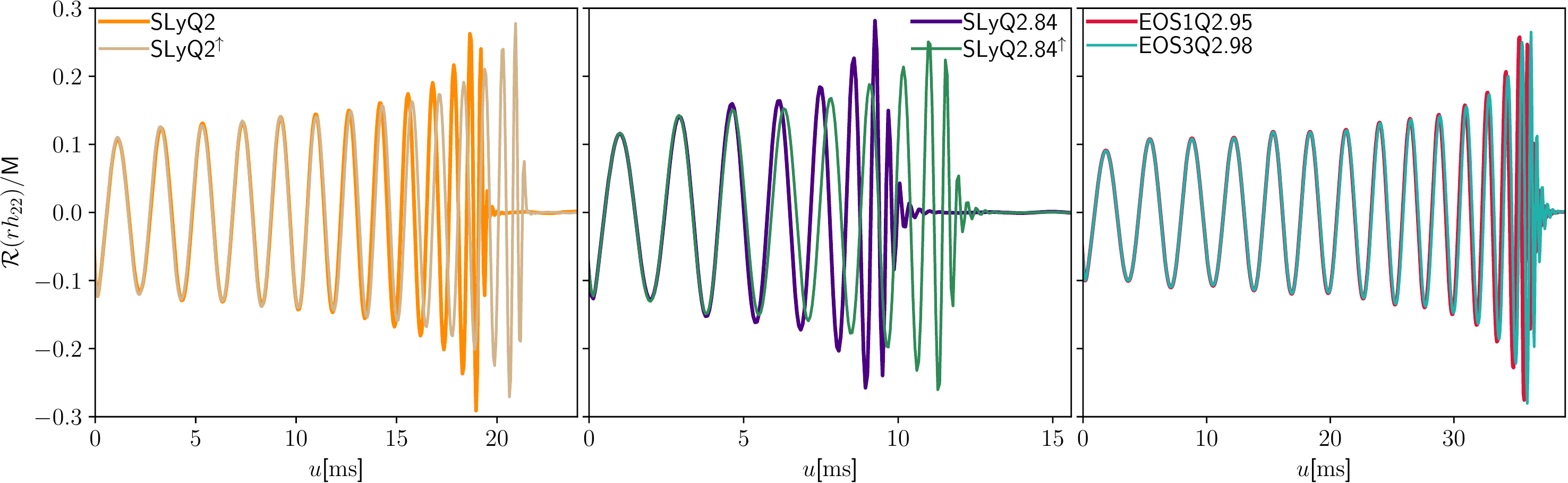}
\caption{(2,2)-mode of the GW strain $rh _{22}$ for six configurations with varying mass ratio, EOS, and spin of the BH (Tab.~\ref{tab:config}). GWs are
abruptly damped when there is tidal disruption of the NS, e.g., SLyQ2 and SLyQ2$^{\uparrow}$ setups whereas BH ringdown is more prominent
for high masses ratio cases where the NS is simply swallowed.}
\label{fig:gws}
\end{figure*}

We also evolve six additional configurations apart from the three configurations
that we use for comparisons and tests. These configurations explore
different mass ratios, EOSs, and BH spins (0 and 0.4). The parameters
are chosen to favor tidal disruption of the NS (within the possibilities
of our ID solver) leading possibly to larger amounts of unbound matter and more massive accretion disks. We would like to note
here that the initial residual eccentricities for all the setups
evolved in this article are~$\mathcal{O}[10^{-3}]$-$\mathcal{O}[10^{-2}]$.
The exact values are listed in Tab.~\ref{tab:config} and are computed using
the trajectories of the compact objects. No eccentricity reduction
procedure has been applied to obtain the IDs. Figure~\ref{fig:postmerger_matter}~shows the evolution of the baryon matter and the BH properties
close to the merger and $\sim 10$ms after the merger for all the setups where the postmerger evolution
is available. In Tab.~\ref{tab:remnant} we list the disk masses ($M^{r>r_\text{AH}}$),
the unbound matter ($M_\text{ej}$), the mass-weighted ejecta velocity ($\bar{v}_{\rm ej}$), and the postmerger BH properties for the different setups for quantitative comparison, cf.\ Sec.~IIC of Ref.~\cite{Chaurasia:2018zhg} and references therein for details.

For the highest mass ratio BHNS system (SLyQ4.76) that we simulate
the unbound matter and disks are negligible ($< 10^{-4} M_\odot$  and $< 10^{-5} M_\odot$, respectively).
This is true in general for high mass ratio BHNS systems where the NS is barely subject to any tidal disruption if the companion BH is nonspinning. Lower mass ratio setups
lead to disks that increase with decreasing mass ratio. The disk mass is
further increased for increasing spin of the BH, i.e., as expected
we find aligned spin BH leads to larger disk mass.
We find a good agreement between the disk mass as reported for \texttt{HBQ2M135}
(see Ref.~\cite{Kyutoku:2010zd}, Tab.~VIII, $M_{\rm disk} = 0.032$) and our HBQ2
setup evolved using \bam.  The trend in the ejected
mass and its average velocity is more complicated and can vary by 50\% among different resolutions.
However, a general trend is again that with spin of the BH the amount of unbound matter increases.

The postmerger BH properties for the \texttt{HBQ2M135} (see Ref.~\cite{Kyutoku:2010zd}, Tab.~VIII,
$M_{\rm BH} = 3.957 M_\odot$ and spin $\chi _{\rm BH} = 0.67$) are in good agreement with our~\bam~evolved
HBQ2 setup. In high mass ratio mergers, where the compact objects are initially irrotational the final spin of the BH is smaller as compared to lower mass ratio setups or setups where the companion BH is initially spinning.
This can be understood as follows: 
In systems with high mass ratio the NS is not much tidally disrupted and therefore
the matter is directly swallowed by the BH. Due to this, the BH is perturbed and undergoes the ringdown phase associated with emission of quasinormal modes. Hence, more GWs are emitted and carry away energy and angular momentum from the system, which leads to a smaller final BH spin. Whereas for lower mass ratios or systems where
the BH is spinning, the NS is tidally disrupted leading 
to an absence of BH ringdown waveforms related
to the BH quasinormal modes in the merger and the ringdown phases, cf. Fig.~\ref{fig:3Dmatter}.
Here the GW amplitude damps abruptly after the inspiral phase when the disrupted material forms 
a relatively low density and nearly axisymmetric matter distribution around the BH, suppressing GW emission. This leads to the final BH to have larger spins as more angular momentum is available to the system.
The waveforms shown in Fig.~\ref{fig:gws}~for the remaining six configurations show the abrupt damping
of the GWs for the cases where the NS is tidally disrupted. Furthermore, they
follow the expected trends; aligned spin systems SLQ2$^{\uparrow}$ and SLyQ3$^{\uparrow}$
have a delayed merger as compared to their non-spinning counterparts SLQ2 and SLyQ3 respectively.
Overall, the disk mass and postmerger BH property estimates are more robust as compared to the ones for unbound matter.

Finally, Fig.~\ref{fig:3Dmatter} shows the 3D time evolution of the bound (top row) 
and the unbound (bottom row) matter for the SLQ2$^{\uparrow}$ configuration.
Due to the comparable masses for a BHNS system (Q$=2$) and the aligned spin of the BH, 
the NS is tidally disrupted before the merger (first column). 
This disruption also causes noticeable ejecta that leaves the system. 
The bound matter then forms a disk surrounding the BH and the unbound
matter, mostly expected to be neutron-rich, expands further (second and third columns). 
In the final stages, the disk still having angular momentum support slowly accretes 
onto the BH while the ejected matter starts to leave the shown part of the computational domain 
(fourth columns). 
Most of the material that is unbound and that will leave the system originates 
from the tidal tail of the NS due to strong torque. Because of this mechanism, the 
ejected material is contained within a small azimuthal angle 
and the ejecta is not distributed axisymmetrically, but as seen in~Fig.~\ref{fig:3Dmatter}, 
there is also clear poloidal dependence of the ejected material. 
Hence, it will be of importance that kilonova models for BHNSs not only 
incorporate a $\theta$ dependence~\cite{Kasen:2017sxr,Wollaeger:2017ahm,Perego:2017wtu,
Bulla:2019muo,Kawaguchi:2019nju,Dietrich:2020lps,Wollaeger:2021qgf}, but also a $\phi$-dependence of the 
ejecta profiles, e.g., using full 3D profiles from numerical-relativity simulations.

\section{Summary}
\label{sec:summary}

In this paper, we presented the first set of BHNS simulations performed with the \bam\ code. 
In total, we evolved nine configurations of which we use three for code comparison and tests. We find that~\bam\ evolutions are in good agreement with \texttt{SACRA} simulations, also using the moving puncture gauge, and with excision simulations done with SpEC by the SXS collaboration.

In our simulations, the \z4c scheme with its constraint damping properties has been essential to be able to evolve the stuffed BH. Stuffing was necessary since we used excision ID while the simulations have been performed with the moving puncture gauge.  
With the \z4c scheme, the overall quality of the simulation seemed good and potentially at the quality that is required for waveform model development. However, we had difficulties in producing adequately convergent initial configurations, hence, a clear quantitative assessment of the numerical uncertainties beyond comparison with previously published data and the simple computation of GW phase differences between different resolutions have not been possible. 
We suspect this to be caused by excision initial data produced with \lorene. 
Therefore, we plan more tests with newer solvers like the \texttt{FUKA} 
solver~\cite{Papenfort:2021hod,FukaCode} in the future.
Finally, we find that for the~\bam~evolved setups the disk mass and the postmerger BH properties are robustly estimated and are consistent with the published literature.

\appendix

\section{GW200115}

\begin{figure}[t]
\centering
\includegraphics[width=0.95\linewidth]{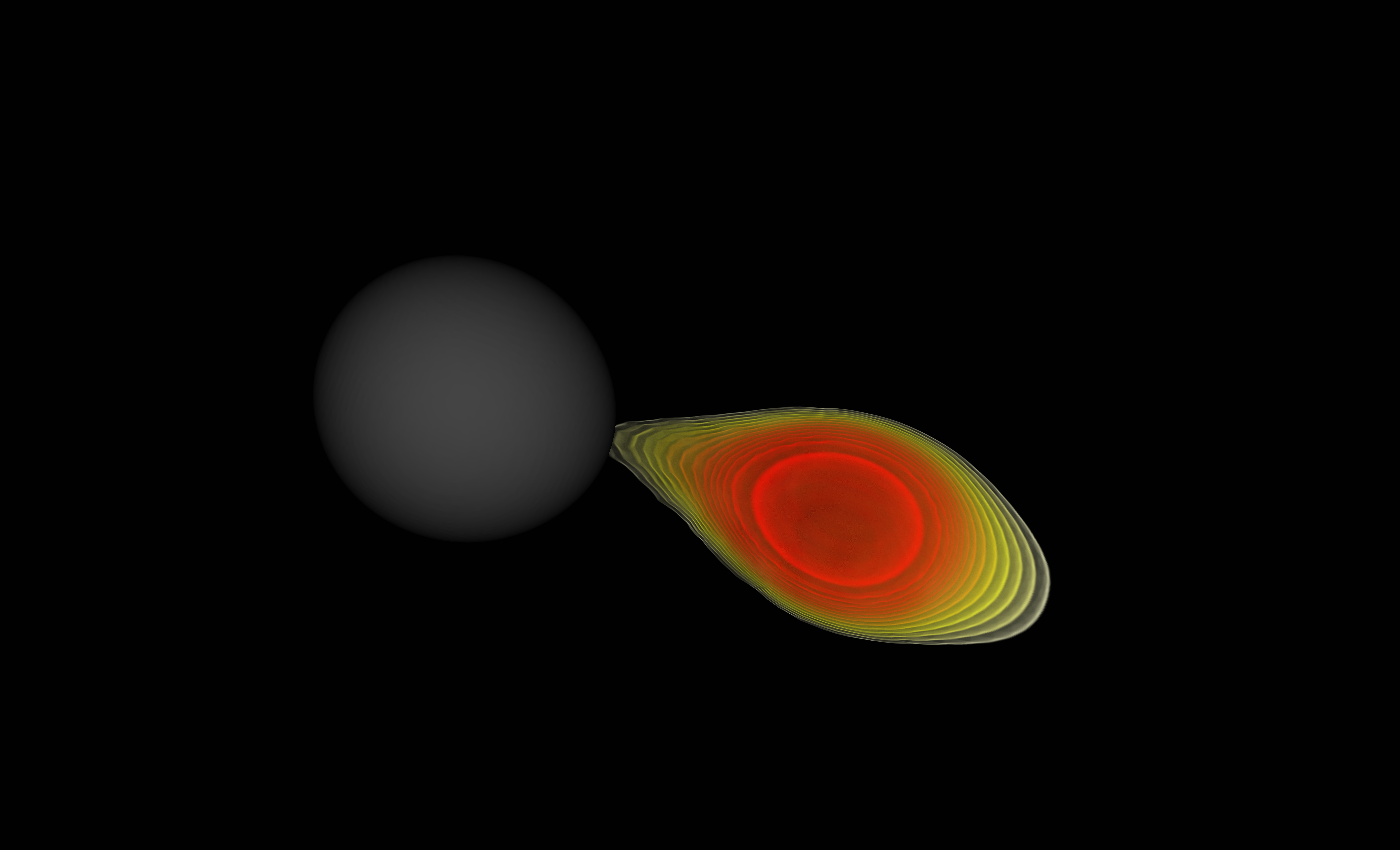}
\includegraphics[width=0.95\linewidth]{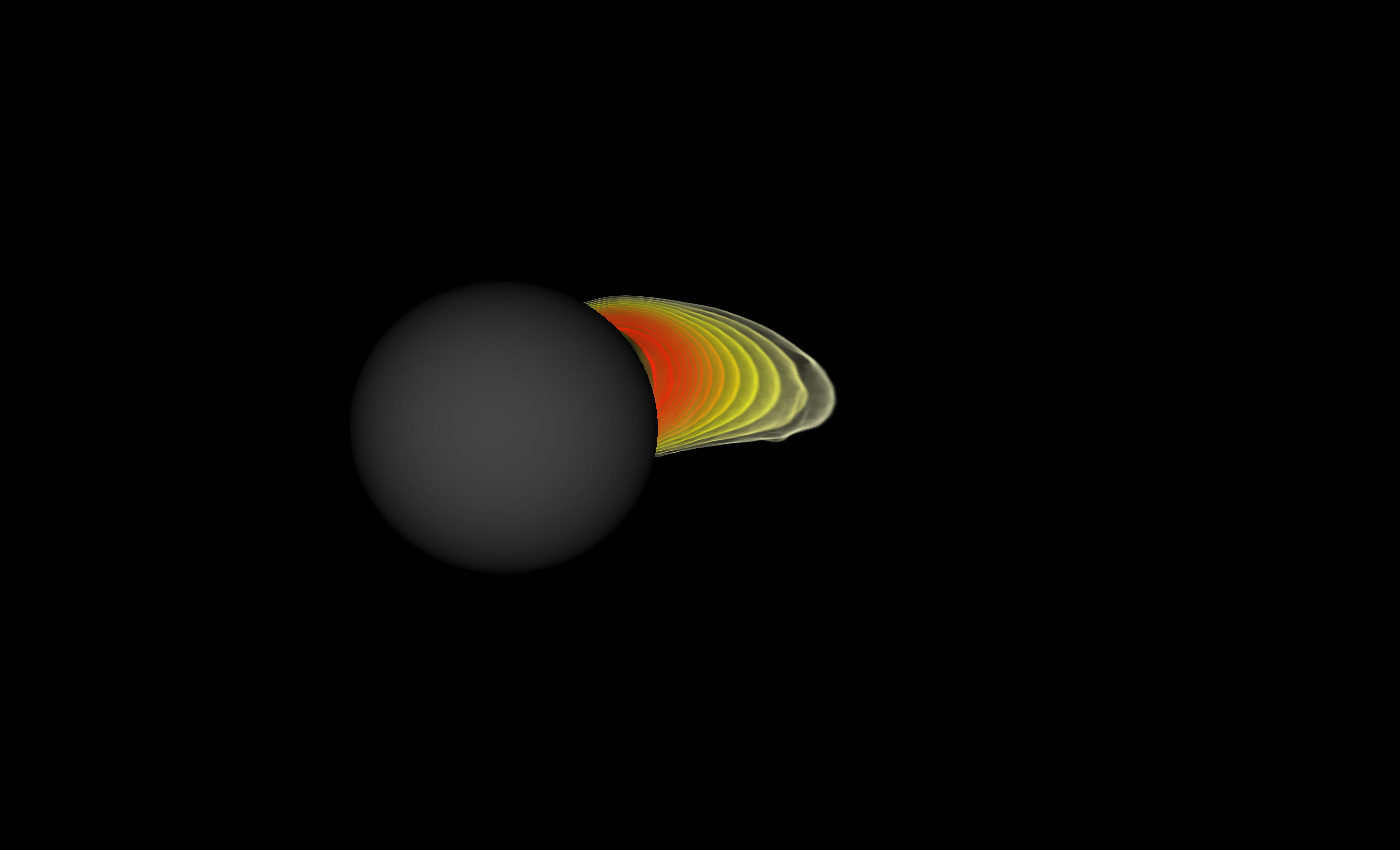}
\includegraphics[width=0.95\linewidth]{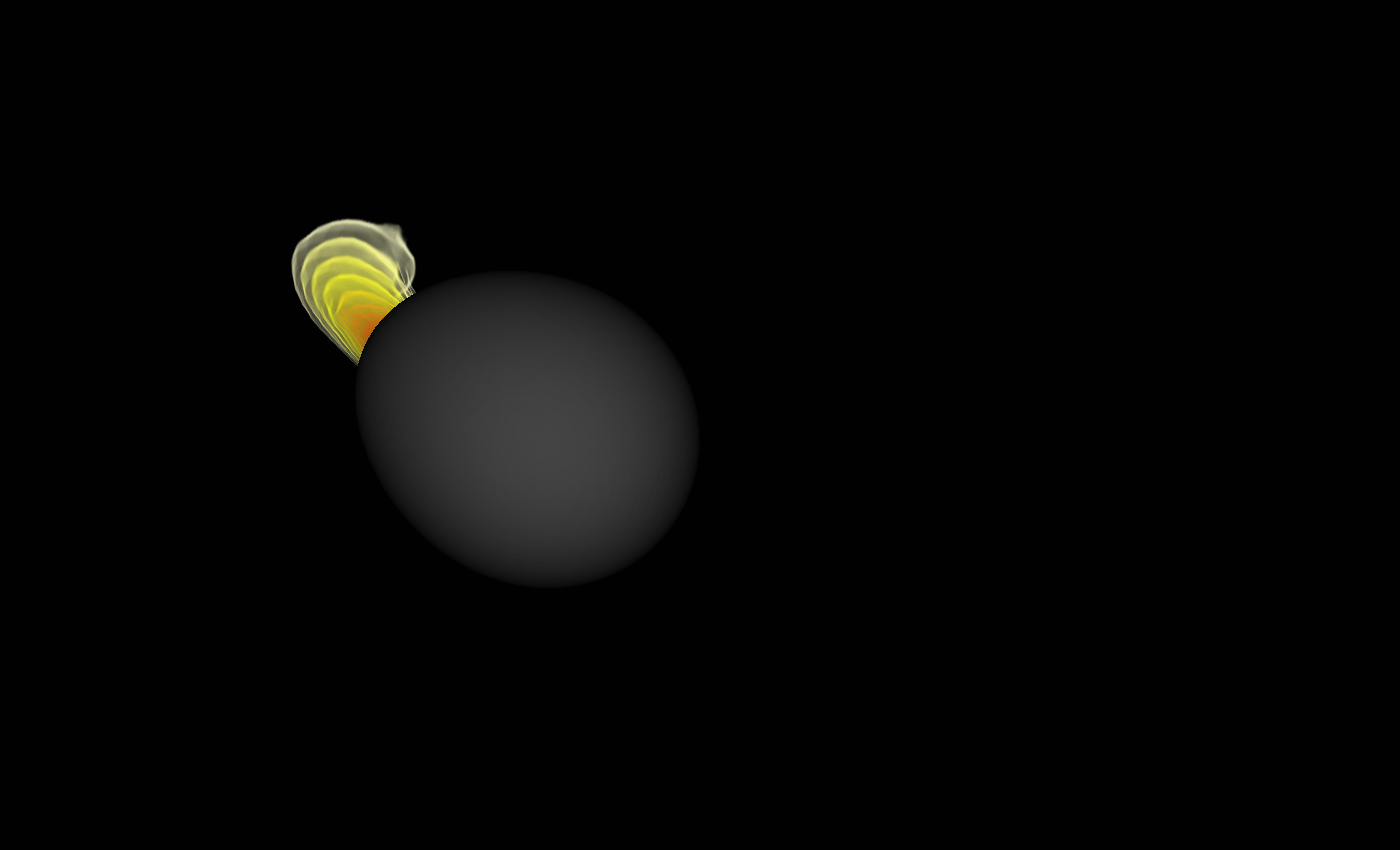}
\caption{3D volume rendering of the rest mass density for our GW200115-like system. 
         We visualize the BH as a contour line of the 
         lapse function with $\alpha=0.37$.}
\label{fig:GW200115}
\end{figure}

While this work was being finalized, the announcement of GW200105 and GW200115~\cite{LIGOScientific:2021qlt}
enhanced further the interest in the simulation of BHNS systems. 
For this purpose and in preparation of visualizations for public outreach, 
we have simulated a system consisting of a non-spinning BH with 
a mass of $6.1M_\odot$ and an irrotational NS with a mass of $1.4M_\odot$ 
described by the SLy EOS. 
These parameters are broadly consistent with the extracted parameters of GW200115. 
We present snapshots of the simulation in Fig.~\ref{fig:GW200115} and refer 
to a full animation including also the GW signal to the material released during the 
announcement of GW200115\footnote{\url{https://www.youtube.com/watch?v=Rd3p3xPtWn4}}. 

As visible in Fig.~\ref{fig:GW200115} and in agreement with Ref.~\cite{LIGOScientific:2021qlt} 
as well as the non-detection of an electromagnetic signal, 
e.g., \cite{Coughlin:2020fwx,Kasliwal:2020wmy,Anand:2020eyg,Paterson:2020mmd}, 
we find that the NS gets swallowed completely by the BH without being tidally disrupted.
Hence, we find no noticeable disk surrounding the final BH ($M_{\rm disk} \lesssim 10^{-5}M_\odot$) 
and no noticeable ejecta material ($M_{\rm ej} \lesssim 5 \times 10^{-4} M_\odot$). 

\begin{acknowledgments}
  We thank Koutarou Kyutoku for providing the waveforms for
  comparison with SACRA and also for very helpful discussions, 
  Philippe Grandcl\'{e}ment for clarifications about
  the~\lorene~BHNS solver, and 
  Nils Fischer, Sergei Ossokine, Harald Pfeiffer for support creating Fig.~\ref{fig:GW200115}.
  We also acknowledge discussions with B.~Br\"ugmann, F.~M.~Fabbri, A.~Rashti, 
  W.~Tichy, M.~Ujevic~Tonino, and F.~Torsello.
  S.~V.~C.~was funded by the research environment grant ``Gravitational Radiation and
  Electromagnetic Astrophysical Transients (GREAT)"
  funded by the Swedish Research council (VR) under Dnr. 2016-06012.
  SR has been supported by the Swedish Research Council (VR) under grant number
  2020-05044, by the Swedish National Space Board under grant number Dnr. 107/16, the research environment grant GREAT and by the Knut and Alice Wallenberg Foundation (KAW 2019.0112). TD acknowledges funding through the Max Planck Society. 
  
  Computations were performed on Beskow at SNIC
  [project numbers SNIC 2020/1-34 and SNIC 2020/3-25],
  on Lise/Emmy of the 
  North German Supercomputing Alliance (HLRN) [project bbp00049], 
  on HAWK at the High-Performance Computing Center Stuttgart 
  (HLRS) [project GWanalysis 44189], 
  on SuperMUC\_NG of the Leibniz Supercomputing Centre (LRZ)
  [project pn29ba], and on the ARA cluster of the University of Jena.
\end{acknowledgments}

\newpage

\bibliography{paper20210719.bbl}

\end{document}